\documentstyle[11pt]{article}
\begin{document}

\begin{flushright}
{hep-ph/0103xxx}\\

\end{flushright}
\vspace{1cm}
\begin{center}
{\Large \bf  Effective Lagrangian for  $\overline{s}bg$ and
$\overline{s}b\gamma$ Vertices\\
\vspace{0.5cm}
in the mSUGRA model}
\vspace{.2cm}
\end{center}
\vspace{1cm}
\begin{center}
{Tai-Fu Feng$^{a,b,d}$\hspace{0.5cm}Xue-Qian Li$^{a,b,d}$}
\hspace{.5cm}Guo-Li Wang$^{a,c,d}$\\
\vspace{.5cm}

{$^a$CCAST (World Laboratory), P.O.Box 8730,
Beijing 100080, P. R. China}\\
{$^b$Department of Physics, Nankai
University, Tianjin 300070, P. R. China}\footnote{Postal address}\\
{$^c$Department of Physics, Fujian Normal
University, Fuzhou, 350007, P. R. China}\\
{$^d$Institute of Theoretical Physics, Academia Sinica, P.O. Box
2735, Beijing 100080, P. R. China}\\
\vspace{.5cm}

\end{center}
\hspace{3in}

\begin{center}
\begin{minipage}{11cm}
{\large\bf Abstract}

{\small Complete expressions of the $\overline{s}bg$ and
$\overline{s}b\gamma$ vertices are derived in the framework
of supersymmetry with minimal flavor violation. With the minimal
supergravity (mSUGRA) model, a numerical analysis of the
supersymmetric contributions to the Wilson Coefficients at the weak scale
is presented.}
\end{minipage}
\end{center}

\vspace{4mm}
{\large{\bf PACS numbers:} 11.30.Er, 12.15.Ff, 12.60.Jv, 13.10+q.} \\

{\large{\bf Keywords:} Supersymmetry, Flavor violation, Effective Lagrangian.}

\vspace{1cm}

\section{Introduction}
The rare $B$ decays serve as a good test for new physics beyond
the standard model (SM) since they are not seriously affected by
the uncertainties due to long distance effects.
The forthcoming B-factories will make more precise measurements on the
rare B-decay processes and those measurements would set more strict
constraints on the new physics beyond  SM.
The main purpose of investigation of B-decays, especially the rare decay
modes is to search for traces of new physics and determine its
parameter space. In all the extensions of SM, the supersymmetry is
considered as one of the most plausible candidates.
In the general supersymmetric extension
of SM, new sources of flavor violation may appear in those
soft breaking terms\cite{Gabrielli}. Applying the mass insertion method,
the influence of those non-universal soft breaking terms on
various flavor changing neutral current (FCNC) processes are
discussed in literatures\cite{collect}. However, too many free
parameters which exist in the supersymmetry model with non-universal soft breaking
terms decrease the model prediction ability. Thus for a practical
calculation whose results can be compared with the data, one needs
to reduce the number of the free parameters in some way, i.e.
by enforcing some physical conditions and assuming reasonable
symmetries. A realization of this idea is the
minimal supergravity (mSUGRA), which is fully specified by only five
parameters\cite{Fengj}. In this work, we perform a strict analysis on the
$\overline{s}bg\;(\overline{s}b\gamma)$ effective Lagrangian
in the minimal flavor violation supersymmetry up to the leading order.
The NLO SUSY-QCD corrections to those processes have been evaluated in
our another work \cite{fengtf}.

The most general form of the
superpotential which does not violate gauge invariance and the
conservation laws in SM is
\begin{eqnarray}
&&{\cal W}=\mu \epsilon_{ij}\hat{H}_{i}^{1}\hat{H}_{j}^{2}+
\epsilon_{ij}h_{l}^{I}\hat{H}_{i}^{1}\hat{L}^{I}_{j}\hat{R}^{I}-
h_{d}^{I}(\hat{H}_{1}^{1}\hat{Q}^{I}_{2}-\hat{H}_2^1V^{IJ}
\hat{Q}_1^J)\hat{D}^{I}-
h_{u}^{I}(\hat{H}_{1}^{2}V^{*JI}\hat{Q}^{J}_{2}-
\hat{H}_2^2\hat{Q}_1^I)\hat{U}^{I}.
\label{superpotential}
\end{eqnarray}
Here $\hat{H}^{1}$, $\hat{H}^{2}$ are Higgs superfields;
$\hat{Q}^{I}$ and $\hat{L}^{I}$ are quark and lepton superfields
in doublets of the weak SU(2) group,
where I=1, 2, 3 are the indices of generations;
the rest superfields
$\hat{U}^{I}$, $\hat{D}^{I}$ and $\hat{R}^{I}$
are quark superfields
of the u- and d-types and  charged leptons
in singlets of the weak SU(2) respectively. Indices i, j are contracted
for the SU(2) group, and $h_{l}$, $h_{u,d}$
are the Yukawa couplings. In order to break the supersymmetry,
the soft breaking terms are introduced as
\begin{eqnarray}
&&{\cal L}_{soft}=-m_{H^1}^2H_i^{1*}H_i^1-m_{H^2}^2H_i^{2*}H_i^2
-m_{L^I}^2\tilde{L}_i^{I*}\tilde{L}_i^{I}\nonumber \\
&&\hspace{2.cm}-m_{R^I}^2\tilde{R}^{I*}\tilde{R}^{I}
-m_{Q^I}^2\tilde{Q}_i^{I*}\tilde{Q}_i^{I}
-m_{U^I}^2\tilde{U}^{I*}\tilde{U}^{I}\nonumber \\
&&\hspace{2.cm}-m_{D^I}^2\tilde{D}^{I*}\tilde{D}^{I}
+(m_1\lambda_B\lambda_1+m_2\lambda_A^i\lambda_A^i\nonumber \\
&&\hspace{2.cm}+m_3\lambda_G^a\lambda_G^a+h.c.)
+\Big[B\mu\epsilon_{ij}H_i^1H_j^2
+\epsilon_{ij}A_l^Ih_{l}^{I}H_{i}^{1}\tilde{L}^{I}_{j}\tilde{R}^{I}
\nonumber \\
&&\hspace{2.cm}-A_d^Ih_{d}^{I}(H_{1}^{1}\tilde{Q}^{I}_{2}
-H_2^1V^{IJ}\tilde{Q}_1^J)\tilde{D}^{I}-
A_u^Ih_{u}^{I}(H_{1}^{2}V^{*JI}\tilde{Q}^{J}_{2}-
H_2^2\tilde{Q}_1^I)\tilde{U}^{I}+h.c.\Big],
\label{soft}
\end{eqnarray}
where $m_{H^1}^2,\;m_{H^2}^2,\;m_{L^I}^2,\;m_{R^I}^2,\;m_{Q^I}^2,\;
m_{U^I}^2$ and $m_{D^I}^2$ are the parameters in unit of mass squared,
$m_3,\;m_2,\;m_1$ denote the masses of $\lambda_G^a\;(a=1,\;2,\;
\cdots\;8),\;\lambda_A^i\;(i=1,\;2,\;3)$ and $\lambda_B$, which are
the $SU(3)\times
SU(2)\times U(1)$ gauginos. $B$ is a free parameter in unit of mass.
$A_{l}^{I},\;A_{u}^{I},\;A_{d}^{I}\;(I=1,\;2,\;3)$ are the soft breaking
parameters that result in mass splitting between leptons, quarks and
their supersymmetric partners. Taking into account of the soft breaking
terms Eq.(\ref{soft}), we can study the
phenomenology within the minimal supersymmetric extension of
the standard model (MSSM). The resultant mass matrix of the up-type scalar
quarks is written as
\begin{eqnarray}
&&m_{\tilde{U}^I}^2=\left(
\begin{array}{cc}
m_{Q^I}^2+m_{u^I}^2+(\frac{1}{2}-\frac{2}{3}\sin^2\theta_{\rm W})
\cos 2\beta m_{\rm Z}^2 & -m_{u^I}(A_u^I+\mu\cot\beta) \\
-m_{u^I}(A_u^I+\mu\cot\beta) & m_{U^I}^2+m_{u^I}^2+
\frac{2}{3}\sin^2\theta_{\rm W}\cos 2\beta m_{\rm Z}^2
\end{array} \right),
\label{stmass}
\end{eqnarray}
and the corresponding mass matrix of the down-type scalar quarks is
\begin{eqnarray}
&&m_{\tilde{D}^I}^2=\left(
\begin{array}{cc}
m_{Q^I}^2+m_{d^I}^2+(\frac{1}{2}+\frac{1}{3}\sin^2\theta_{\rm W})
\cos 2\beta m_{\rm Z}^2 & -m_{d^I}(A_d^I+\mu\tan\beta) \\
-m_{d^I}(A_d^I+\mu\tan\beta) & m_{D^I}^2+m_{d^I}^2-
\frac{1}{3}\sin^2\theta_{\rm W}\cos 2\beta m_{\rm Z}^2
\end{array} \right),
\label{sbmass}
\end{eqnarray}
with $m_{u^I},\;m_{d^I}\;(I=1,\;2,\;3)$ being the masses of
the I-th generation quarks.
One difference between the MSSM and
SM is the Higgs sector. There are four charged scalars,
two of them are physical massive Higgs bosons and other are
massless Goldstones in the SUSY extension. The mixing matrix can be written as:
\begin{equation}
{\cal Z}_{H}=\left(
\begin{array}{cc}
\sin\beta & -\cos\beta \\
\cos\beta & \sin\beta \end{array}
\right) \label{zh} \end{equation}
with $\tan\beta=\frac{\upsilon_{2}}{\upsilon_{1}}$ and $\upsilon_1,\upsilon_2$
being the vacuum-expectation values of the two Higgs scalars.
Another matrix that we will use in the later derivation is the chargino mixing
matrix. The SUSY partners of the charged Higgs and $W^{\pm}$
combine to give four Dirac fermions: $\chi^{\pm}_{1}$,
$\chi^{\pm}_{2}$. The two mixing matrices ${\cal Z}^{\pm}$
appearing in the Lagrangian are defined as
\begin{eqnarray}
&&({\cal Z}^{-})^{T}{\cal M}_{c}{\cal Z}^{+} = diag(m_{_{\chi_{1}}},
m_{_{\chi_{2}}}),
\label{zpm}
\end{eqnarray}
where ${\cal M}_{c}$ is the mass matrix of charginos.
In a similar way, $Z_{U,D}$ diagonalize the mass
matrices of the up- and down-type squarks respectively:
\begin{eqnarray}
&&{\cal Z}_{_{{\tilde U}^I}}^{\dag}m^2_{_{{\tilde U}^I}}
{\cal Z}_{_{{\tilde U}^I}}=
diag(m^2_{_{{\tilde U}^I_1}},\; m^2_{_{{\tilde U}^I_2}})\;,
\nonumber \\
&&{\cal Z}_{_{{\tilde D}^I}}^{\dag}m^2_{_{{\tilde D}^I}}
{\cal Z}_{_{{\tilde D}^I}}=
diag(m^2_{_{{\tilde D}^I_1}},\; m^2_{_{{\tilde D}^I_2}})\;.
\label{zud}
\end{eqnarray}
In the framework of minimal supergravity (mSUGRA), the unification assumptions
at the GUT scale are expressed as\cite{valle}
\begin{eqnarray}
&&A_l^I=A_d^I=A_u^I=A_0\;,\\ \nonumber
&&B=A_0-1\;, \\ \nonumber
&&m_{H^1}^2=m_{H^2}^2=m_{L^I}^2=m_{R^I}^2=m_{Q^I}^2=m_{U^I}^2=
m_{D^I}^2=m_0^2\;,\\ \nonumber
&&m_1=m_2=m_3=m_{\frac{1}{2}}\;.
\label{unifi}
\end{eqnarray}
Under these assumptions, the mSUGRA is specified by five parameters:
$$A_0,\; m_0,\; m_\frac{1}{2},\;\tan\beta,\; sgn(\mu),$$
and the flavor structure of the model is similar to SM, i.e.
flavors change only via the CKM matrix.

The supersymmetric contributions will modify the Wilson
Coefficients of the effective $\overline{s}bg$ and $\overline{s}b\gamma$
vertices.
For the W-boson propagator, we adopt the nonlinear $R_\xi$ gauge
whose gauge fixing term is \cite{weinberg}
\begin{equation}
{\cal L}_{_{gauge-fixing}}=-\frac{1}{\xi}f^\dagger f
\label{gaugefix}
\end{equation}
with $f=(\partial_\mu W^{+\mu}-ieA_\mu W^{+\mu}-i\xi m_{_{\rm W}}
\phi^+)$
in our calculations. A thorough discussion about the gauge invariance
was given by Deshpande et al. \cite{Deshpande1,Deshpande2}.

As in the case of SM \cite{Grigjanis}, the operator basis for $b\rightarrow
sg$ in the supersymmetry consists of
\begin{eqnarray}
&&{\cal O}_1=\frac{1}{(4\pi)^2}\bar{s}(i\;/\!\!\!\!\!D)^3\omega_-b\;,
\nonumber \\
&&{\cal O}_2=\frac{1}{(4\pi)^2}\bar{s}\{i\;/\!\!\!\!\!D,
g_sG\cdot\sigma\}\omega_-b\;,
\nonumber \\
&&{\cal O}_3=\frac{1}{(4\pi)^2}\bar{s}iD_{\mu}(ig_sG^{\mu\nu})\gamma_{\nu}
\omega_-b\;,
\nonumber \\
&&{\cal O}_4=\frac{1}{(4\pi)^2}\bar{s}(i\;/\!\!\!\!\!D)^2
(m_s\omega_-+m_b\omega_+)b\;,
\nonumber \\
&&{\cal O}_5=\frac{1}{(4\pi)^2}\bar{s}g_sG\cdot \sigma(m_s\omega_-
+m_b\omega_+)b\;.
\label{basbsg}
\end{eqnarray}
In these operators, $D_{\mu}\equiv \partial_{\mu}-ig_sG_{\mu}$
and
$G_{\mu\nu}\equiv G_{\mu\nu}^aT^a$ denotes the gluon field strength
tensor with $G_{\mu\nu}^a=\partial_{\mu}G_{\nu}^a-\partial_{\nu}
G_{\mu}^a+g_sf^{abc}G_{\mu}^bG_{\nu}^c$, and $G\cdot \sigma\equiv
G_{\mu\nu}\sigma^{\mu\nu}$.

For transition $b\rightarrow s\gamma$, the operator basis is
somewhat different from those in eq.(\ref{basbsg}) and the changes
are reflected in the following replacements:
\begin{eqnarray}
&&{\cal O}_2\rightarrow {\cal O}_6=\frac{1}{(4\pi)^2}\bar{s}\{i\;/\!\!\!\!\!D,
eQ_dF\cdot\sigma\}\omega_-b\;,
\nonumber \\
&&{\cal O}_3\rightarrow {\cal O}_7=\frac{1}{(4\pi)^2}\bar{s}iD_{\mu}
(ieQ_dF^{\mu\nu})\gamma_{\nu}\omega_-b\;,
\nonumber \\
&&{\cal O}_5\rightarrow {\cal O}_8=\frac{1}{(4\pi)^2}\bar{s}eQ_dF\cdot
\sigma(m_s\omega_-+m_b\omega_+)b
\label{basgamma}
\end{eqnarray}
with $F_{\mu\nu}$ being the electromagnetic field strength tensor and
$F\cdot \sigma\equiv F_{\mu\nu}\sigma^{\mu\nu}$.

\section{The effective Lagrangian for $\bar{s}bg\;(\bar{s}b\gamma)$}

\begin{center}
\begin{figure}
\setlength{\unitlength}{1mm}
\begin{picture}(230,200)(55,90)
\put(50,30){\includegraphics{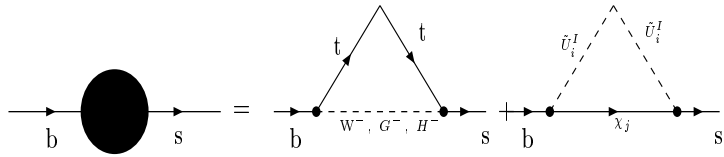}}
\end{picture}
\caption[]{The one-loop self-energy diagrams for $b\rightarrow s$
in the SUSY model with minimal flavor violation }
\label{fig1}
\end{figure}
\end{center}

At first, we present the analysis of $\bar{s}b$-mixing. The self-energy
diagrams are drawn in Fig.\ref{fig1}. The unrenormalized $\bar{s}b$
self-energy is given as
\begin{eqnarray}
&&\Sigma=\frac{ig_2^2}{32\pi^2}\sum\limits_{i=u,c,t}V_{ib}V_{is}^*
\Big\{\Big(A_0(x_i,x_{_H},x_{_{\tilde{U}_\alpha^i}},x_{_{\chi_\beta}})
+\frac{p^2}{m_{_{\rm W}}^2}A_1(x_i,x_{_H},x_{_{\tilde{U}_\alpha^i}},
x_{_{\chi_\beta}})\Big)/\!\!\! p\omega_- \nonumber \\
&&\hspace{1.cm}+ \Big(B_0(x_i,x_{_H},
x_{_{\tilde{U}_\alpha^i}},x_{_{\chi_\beta}})
+\frac{p^2}{m_{_{\rm W}}^2}B_1(x_i,x_{_H},x_{_{\tilde{U}_\alpha^i}},
x_{_{\chi_\beta}})\Big)(m_s\omega_-+m_b\omega_+) \nonumber \\
&&\hspace{1.cm}+C_0(x_i,x_{_H},x_{_{\tilde{U}_\alpha^i}},x_{_{\chi_\beta}})
\frac{m_bm_s}{m_{_{\rm W}}^2}/\!\!\! p\omega_+\Big\}
\label{self0}
\end{eqnarray}
with the symbolic definitions  $x_i=\frac{m_i^2}{m_{_{\rm W}}^2},
x_{_H}=\frac{m_{_{H^+}}^2}{m_{_{\rm W}}^2},x_{_{\tilde{U}_\alpha^i}}
=\frac{m_{_{\tilde{U}_\alpha^i}}^2}{m_{_{\rm W}}^2},
x_{_{\chi_\beta}}=\frac{m_{_{\chi_\beta}}^2}{m_{_{\rm W}}^2}$
with $i=u\;,c\;,t$.
Those form factors $A_0,A_1,B_0,B_1$ and $C_0$
are complicated functions of the parameters and their explicit
expressions are collected in Appendix.\ref{app1}.

We renormalize the $\bar{s}b$ self-energy according to the well-known
prescription, namely by demanding that the renormalized self-energy $\hat{\Sigma}$
vanishes when one of the external legs is on its mass-shell\cite{Shabalin,
Nanopoulos,Chia}. Obviously, this is a necessary physical condition
which must be satisfied. This is realized as
\begin{eqnarray}
&&\hat{\Sigma}=\frac{ig_2^2}{32\pi^2}\sum\limits_{i=u,c,t}V_{ib}V_{is}^*
\Big\{\Big(A^*+A_0(x_i,x_{_H},x_{_{\tilde{U}_\alpha^i}},x_{_{\chi_\beta}})
+\frac{p^2}{m_{_{\rm W}}^2}A_1(x_i,x_{_H},x_{_{\tilde{U}_\alpha^i}},
x_{_{\chi_\beta}})\Big)/\!\!\! p\omega_- \nonumber \\
&&\hspace{1.cm}+ \Big(B_s^*+B_0(x_i,x_{_H},
x_{_{\tilde{U}_\alpha^i}},x_{_{\chi_\beta}})
+\frac{p^2}{m_{_{\rm W}}^2}B_1(x_i,x_{_H},x_{_{\tilde{U}_\alpha^i}},
x_{_{\chi_\beta}})\Big)m_s\omega_-\nonumber \\
&&\hspace{1.cm}+ \Big(B_b^*+B_0(x_i,x_{_H},
x_{_{\tilde{U}_\alpha^i}},x_{_{\chi_\beta}})
+\frac{p^2}{m_{_{\rm W}}^2}B_1(x_i,x_{_H},x_{_{\tilde{U}_\alpha^i}},
x_{_{\chi_\beta}})\Big)m_b\omega_+\nonumber \\
&&\hspace{1.cm}+\Big(C^*+C_0(x_i,x_{_H},x_{_{\tilde{U}_\alpha^i}},
x_{_{\chi_\beta}})\Big)\frac{m_bm_s}{m_{_{\rm W}}^2}/\!\!\! p\omega_+\Big\}
\;,
\label{selfr}
\end{eqnarray}
where
\begin{eqnarray}
&&A^*=-A_0(x_i,x_{_H},x_{_{\tilde{U}_\alpha^i}},x_{_{\chi_\beta}})
-\frac{m_b^2+m_s^2}{m_{_{\rm W}}^2}\Big(
A_1(x_i,x_{_H},x_{_{\tilde{U}_\alpha^i}},x_{_{\chi_\beta}})+
B_1(x_i,x_{_H},x_{_{\tilde{U}_\alpha^i}},x_{_{\chi_\beta}})\Big)\;,
\nonumber \\
&&B_b^*=-B_0(x_i,x_{_H},x_{_{\tilde{U}_\alpha^i}},x_{_{\chi_\beta}})
-\frac{m_s^2}{m_{_{\rm W}}^2}\Big(
A_1(x_i,x_{_H},x_{_{\tilde{U}_\alpha^i}},x_{_{\chi_\beta}})+
B_1(x_i,x_{_H},x_{_{\tilde{U}_\alpha^i}},x_{_{\chi_\beta}})\Big)\;,
\nonumber \\
&&B_s^*=-B_0(x_i,x_{_H},x_{_{\tilde{U}_\alpha^i}},x_{_{\chi_\beta}})
-\frac{m_b^2}{m_{_{\rm W}}^2}\Big(
A_1(x_i,x_{_H},x_{_{\tilde{U}_\alpha^i}},x_{_{\chi_\beta}})+
B_1(x_i,x_{_H},x_{_{\tilde{U}_\alpha^i}},x_{_{\chi_\beta}})\Big)\;,
\nonumber \\
&&C^*=-\frac{m_bm_s}{m_{_{\rm W}}^2}\Big(A_1(x_i,x_{_H},
x_{_{\tilde{U}_\alpha^i}},x_{_{\chi_\beta}})+
2B_1(x_i,x_{_H},x_{_{\tilde{U}_\alpha^i}},x_{_{\chi_\beta}})+
C_0(x_i,x_{_H},x_{_{\tilde{U}_\alpha^i}},x_{_{\chi_\beta}})\Big)\;.
\label{exp2}
\end{eqnarray}
After carrying out the renormalization procedure described above,
the self-energy is written as
\begin{eqnarray}
&&\hat{\Sigma}=\frac{ig_2^2}{32\pi^2}\sum\limits_{i=u,c,t}V_{ib}V_{is}^*\Big\{
\Big[\frac{p^2-m_b^2-m_s^2}{m_{_{\rm W}}^2}A_1
(x_i,x_{_H},x_{_{\tilde{U}_\alpha^i}},x_{_{\chi_\beta}})-
\frac{m_b^2+m_s^2}{m_{_{\rm W}}^2}
B_1(x_i,x_{_H},x_{_{\tilde{U}_\alpha^i}},x_{_{\chi_\beta}})\Big]
/\!\!\! p\omega_- \nonumber \\
&&\hspace{1.2cm}+\Big[\frac{p^2}{m_{_{\rm W}}^2}B_1(
x_i,x_{_H},x_{_{\tilde{U}_\alpha^i}},x_{_{\chi_\beta}})
+\frac{m_b^2}{m_{_{\rm W}}^2}\Big(A_1(
x_i,x_{_H},x_{_{\tilde{U}_\alpha^i}},x_{_{\chi_\beta}})
+B_1(x_i,x_{_H},x_{_{\tilde{U}_\alpha^i}},x_{_{\chi_\beta}})
\Big)\Big]m_s\omega_- \nonumber \\
&&\hspace{1.2cm}+\Big[\frac{p^2}{m_{_{\rm W}}^2}B_1(
x_i,x_{_H},x_{_{\tilde{U}_\alpha^i}},x_{_{\chi_\beta}})
+\frac{m_s^2}{m_{_{\rm W}}^2}\Big(A_1(
x_i,x_{_H},x_{_{\tilde{U}_\alpha^i}},x_{_{\chi_\beta}})
+B_1(x_i,x_{_H},x_{_{\tilde{U}_\alpha^i}},x_{_{\chi_\beta}})
\Big)\Big]m_b\omega_+ \nonumber \\
&&\hspace{1.2cm}-\frac{m_bm_s}{m_{_{\rm W}}^2}\Big(
A_1(x_i,x_{_H},x_{_{\tilde{U}_\alpha^i}},x_{_{\chi_\beta}})
+2B_1(x_i,x_{_H},x_{_{\tilde{U}_\alpha^i}},x_{_{\chi_\beta}})\Big)
/\!\!\! p\omega_+ \Big\}.
\label{selfr1}
\end{eqnarray}
This procedure is exactly the same as that adopted in the SM case\cite{Chia1}.

\begin{center}
\begin{figure}
\setlength{\unitlength}{1mm}
\begin{picture}(230,200)(55,90)
\put(50,30){\includegraphics{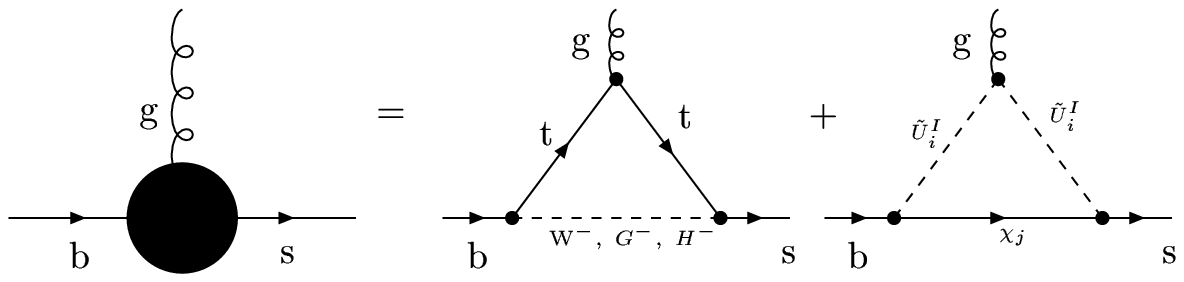}}
\end{picture}
\caption[]{The one-loop diagrams  for $b\rightarrow sg$
in the SUSY model with  minimal flavor violation}
\label{fig2}
\end{figure}
\end{center}

Next, let us calculate the unrenormalized $\bar{s}bg$ vertex $\Gamma_\rho
(p,q)$ corresponding to Fig.\ref{fig2}. Keeping terms up to order
$\frac{p^2,\;q^2}{m_{_{\rm W}}^2}$\cite{Inami,Grigjanis,Grinstein,Cella,
Misiak}, we have
\begin{eqnarray}
&&\Gamma_\rho^{^{b\rightarrow sg}}=g_sT^a\frac{ig_2^2}{32\pi^2}\sum
\limits_{i=u,c,t}V_{ib}V_{is}^*\Big\{A_0(x_i,x_{_H},x_{_{\tilde{U}_\alpha^i}},
x_{_{\chi_\beta}})\gamma_\rho\omega_-
\nonumber \\
&&\hspace{1.2cm}+A_1(x_i,x_{_H},x_{_{\tilde{U}_\alpha^i}},
x_{_{\chi_\beta}})\frac{p^2\gamma_\rho+(p+q)^2\gamma_\rho
+/\!\!\!p\gamma_\rho /\!\!\!p}{m_{_{\rm W}}^2}\omega_-
+F_1(x_i,x_{_H},x_{_{\tilde{U}_\alpha^i}},x_{_{\chi_\beta}})
\frac{q^2}{m_{_{\rm W}}^2}\gamma_\rho\omega_-
\nonumber \\
&&\hspace{1.2cm}+F_2(x_i,x_{_H},x_{_{\tilde{U}_\alpha^i}},x_{_{\chi_\beta}})
\frac{/\!\!\!p\gamma_\rho /\!\!\!q}{m_{_{\rm W}}^2}\omega_-
+F_3(x_i,x_{_H},x_{_{\tilde{U}_\alpha^i}},x_{_{\chi_\beta}})
\frac{/\!\!\!q\gamma_\rho /\!\!\!p}{m_{_{\rm W}}^2}\omega_-
\nonumber \\
&&\hspace{1.2cm}+F_4(x_i,x_{_H},x_{_{\tilde{U}_\alpha^i}},x_{_{\chi_\beta}})
\frac{/\!\!\!q\gamma_\rho /\!\!\!q}{m_{_{\rm W}}^2}\omega_-
+B_1(x_i,x_{_H},x_{_{\tilde{U}_\alpha^i}},x_{_{\chi_\beta}})
\frac{1}{m_{_{\rm W}}^2}\Big((/\!\!\!p+/\!\!\!q)\gamma_\rho
+\gamma_\rho/\!\!\!p\Big)(m_s\omega_-+m_b\omega_+)
\nonumber \\
&&\hspace{1.2cm}
+F_5(x_i,x_{_H},x_{_{\tilde{U}_\alpha^i}},x_{_{\chi_\beta}})
\frac{1}{m_{_{\rm W}}^2}[/\!\!\!q,\gamma_\rho](m_s\omega_-+m_b\omega_+)+
C_0(x_i,x_{_H},x_{_{\tilde{U}_\alpha^i}},x_{_{\chi_\beta}})
\frac{m_bm_s}{m_{_{\rm W}}^2}\gamma_\rho\omega_+\Big\}\;,
\label{bsg0}
\end{eqnarray}
where  $F_i(x_i,x_{_H},x_{_{\tilde{U}_\alpha^i}},x_{_{\chi_\beta}})
\;(i=1,\;\cdots,\;5)$ are collected in Appendix.\ref{app1}.
From  Eq.\ref{self0} and Eq.\ref{bsg0}, it is easy to show that
$\Gamma_\rho^{^{b\rightarrow sg}}$ obeys the Ward-Takahashi identity
\begin{eqnarray}
&&q^\rho \Gamma_\rho^{^{b\rightarrow sg}}(p,q)=
g_sT^a\Big[\Sigma(p+q)-\Sigma(p)\Big]\;.
\label{ward0}
\end{eqnarray}

According to the general principle of renormalization,  $\bar
sbg-$vertex does not exist in the fundamental Lagrangian, thus it
does not need to be renormalized. In other words, the divergence
would be canceled as the physical conditions are taken into
account.
In the nonlinear $R_{\xi}$ gauge, as
well as in the unitary gauge,  the one-loop penguin diagram
results in a divergence. On other side, all the one-loop
diagrams which contribute to the $b\rightarrow sg$ or
$b\rightarrow s\gamma$ processes constitute a convergent subgroup.
Thus obviously, the renormalizations of the penguin and
flavor-changing self energies are associated.
In fact, the Ward-Takahashi identity holds at the
unrenormalized penguin,
to renormalize the $\bar{s}bg$ vertex, we demand that the Ward-Takahashi
identity be preserved by the renormalized vertex $\hat{\Gamma}
_\rho^{^{b\rightarrow sg}}$\cite{Chia1},
\begin{eqnarray}
&&q^\rho \hat{\Gamma}_\rho^{^{b\rightarrow sg}}(p,q)=
g_sT^a\Big[\hat{\Sigma}(p+q)-\hat{\Sigma}(p)\Big]\;.
\label{wardr}
\end{eqnarray}
It is noted that with this requirement,
just as in the SM case \cite{Chia1}, the renormalization
of the $\bar sbg$ vertex is realized when we renormalize the
self-energy by enforcing the physical condition $\hat{\Sigma}=0$
as one of the external legs being on its mass shell. Moreover,
indeed, the renormalization scheme of the $\bar{s}bg$ vertex
pledges the current conservation for an on-shell transition,
since the renormalized self-energies $\hat{\Sigma}(p+q)$ and
$\hat{\Sigma}(p)$ are zero as both $b$ and $s$ are on mass shell
\cite{Chia1}.

This renormalization scheme can be understood from another angle.
The requirement that the Ward-Takahashi identity holds and
condition $\hat{\Sigma}$(on-shell)=0 realize the renormalization
of the $\bar sbg$ vertex and the scheme is equivalent to summing
up the contributions of penguin and
flavor-changing self-energies to the transition $\bar sbg$
at one-loop level. This procedure
can be generalized to two-loop calculations.

Applying Eq.\ref{wardr}, we have
\begin{eqnarray}
&&\hat{\Gamma}_\rho^{^{b\rightarrow sg}}=g_sT^a\frac{ig_2^2}{32\pi^2}\sum
\limits_{i=u,c,t}V_{ib}V_{is}^*\Big\{-\frac{m_b^2+m_s^2}{m_{_{\rm W}}^2}
\Big(A_1(x_i,x_{_H},x_{_{\tilde{U}_\alpha^i}},
x_{_{\chi_\beta}})+
B_1(x_i,x_{_H},x_{_{\tilde{U}_\alpha^i}},
x_{_{\chi_\beta}})\Big)\gamma_\rho\omega_-
\nonumber \\
&&\hspace{1.2cm}+A_1(x_i,x_{_H},x_{_{\tilde{U}_\alpha^i}},
x_{_{\chi_\beta}})\frac{p^2\gamma_\rho+(p+q)^2\gamma_\rho
+/\!\!\!p\gamma_\rho /\!\!\!p}{m_{_{\rm W}}^2}\omega_-
+F_1(x_i,x_{_H},x_{_{\tilde{U}_\alpha^i}},x_{_{\chi_\beta}})
\frac{q^2}{m_{_{\rm W}}^2}\gamma_\rho\omega_-
\nonumber \\
&&\hspace{1.2cm}+F_2(x_i,x_{_H},x_{_{\tilde{U}_\alpha^i}},x_{_{\chi_\beta}})
\frac{/\!\!\!p\gamma_\rho /\!\!\!q}{m_{_{\rm W}}^2}\omega_-
+F_3(x_i,x_{_H},x_{_{\tilde{U}_\alpha^i}},x_{_{\chi_\beta}})
\frac{/\!\!\!q\gamma_\rho /\!\!\!p}{m_{_{\rm W}}^2}\omega_-
\nonumber \\
&&\hspace{1.2cm}+F_4(x_i,x_{_H},x_{_{\tilde{U}_\alpha^i}},x_{_{\chi_\beta}})
\frac{/\!\!\!q\gamma_\rho /\!\!\!q}{m_{_{\rm W}}^2}\omega_-
+B_1(x_i,x_{_H},x_{_{\tilde{U}_\alpha^i}},x_{_{\chi_\beta}})
\frac{1}{m_{_{\rm W}}^2}\Big((/\!\!\!p+/\!\!\!q)\gamma_\rho
+\gamma_\rho/\!\!\!p\Big)(m_s\omega_-+m_b\omega_+)
\nonumber \\
&&\hspace{1.2cm}
+F_5(x_i,x_{_H},x_{_{\tilde{U}_\alpha^i}},x_{_{\chi_\beta}})
\frac{1}{m_{_{\rm W}}^2}[/\!\!\!q,\gamma_\rho](m_s\omega_-+m_b\omega_+)
\nonumber\\
&&\hspace{1.2cm}
-\Big(A_1(x_i,x_{_H},x_{_{\tilde{U}_\alpha^i}},x_{_{\chi_\beta}})
+2B_1(x_i,x_{_H},x_{_{\tilde{U}_\alpha^i}},x_{_{\chi_\beta}})\Big)
\frac{m_bm_s}{m_{_{\rm W}}^2}\gamma_\rho\omega_+\Big\}.
\label{bsgr}
\end{eqnarray}
The terms of dimension-four which are related to the
$\bar{s}\gamma_\rho\omega_\pm b$ vertex  cancel each other as
long as we let $b$ and $s$ quarks be on their mass shells\cite{Grigjanis},
so that we do not need to consider them at all.
We ignore all terms which vanish
as $\frac{m_{u,c}^2}{m_{_{\rm W}}^2}
\rightarrow 0$, whereas keep the part in the coefficients which are proportional to
$\ln{m_{u,c}^2\over m_{_{\rm W}}^2}$ in the final effective
vertex for $b\rightarrow sg$, we can recast  Eq.\ref{bsgr} to a form
with the operator basis given in Eq.\ref{basbsg}:
\begin{equation}
\hat{\Gamma}_\rho^{^{b\rightarrow sg}}=
\frac{4G_F}{\sqrt{2}}\Big\{V_{tb}V_{ts}^*\sum
\limits_{i=1}^5C_i(\mu_{_{\rm W}}){\cal
O}_i+\Big(\frac{4}{3}V_{cb}V_{cs}^*\ln x_c+
\frac{4}{3}V_{ub}V_{us}^*\ln x_u\Big){\cal O}_3\Big\}\;.
\label{bsglag}
\end{equation}
After matching between the
effective theory and the full theory\cite{buras}, we have the
effective Lagrangian for $b\rightarrow sg$ at the weak scale
in the minimal flavor violating supersymmetry as:
\begin{equation}
{\cal L}_{b\rightarrow sg}=
\frac{4G_F}{\sqrt{2}}V_{tb}V_{ts}^*\sum\limits_{i=1}^5
C_i(\mu_{_{\rm W}}){\cal O}_i
\label{bsglag1}
\end{equation}
where
\begin{eqnarray}
&&C_1(\mu_{_{\rm W}})=-\Big[
\frac{5x_t+1}{2(1-x_t)^3}+\frac{x_t^3+2x_t^2}{(1-x_t)^4}\ln x_t\Big]
+\frac{1}{\tan^2\beta}\Big[\frac{x_t^3x_{_H}(\ln x_t-\ln x_{_H}}
{(x_{_H}-x_t)^4}\nonumber \\
&&\hspace{2.cm}+\frac{2x_t^3+5x_t^2x_{_H}-x_tx_{_H}^2}{6(x_{_H}-x_t)^3}\Big]
+2\sum\limits_{\alpha,\;\beta}({\cal A}_3^{^{\alpha,\beta}})^2
\Big[-\frac{x_{_{\chi_\beta}}^2
x_{_{\tilde{U}^3_\alpha}}(\ln x_{_{\tilde{U}^3_\alpha}}
-\ln x_{_{\chi_\beta}})}{(x_{_{\tilde{U}^3_\alpha}}-x_{_{\chi_\beta}})^4}
\nonumber \\
&&\hspace{2.cm}-\frac{2x_{_{\chi_\beta}}^2+5x_{_{\chi_\beta}}
x_{_{\tilde{U}^3_\alpha}}-
x_{_{\tilde{U}^3_\alpha}}^2}{6(x_{_{\chi_\beta}}
-x_{_{\tilde{U}^3_\alpha}})^3}\Big]\;,
\nonumber \\
&&C_2(\mu_{_{\rm W}})=x_t\Big[
\frac{5x_t-2}{4(1-x_t)^4}\ln x_t+\frac{-4+13x_t-3x_t^2}{8(1-x_t)^3}\Big]
+\frac{1}{\tan^2\beta}\Big[-\frac{x_t^2x_{_H}^2(\ln x_t-\ln x_{_H})}
{4(x_{_H}-x_t)^4}
\nonumber \\
&&\hspace{2.cm}-\frac{2x_tx_{_H}^2+5x_t^2x_{_H}-x_t^3}{24(x_{_H}-x_t)^3}\Big]
+\sum\limits_{\alpha,\;\beta}({\cal A}_3^{^{\alpha,\beta}})^2
\Big[\frac{x_{_{\chi_\beta}}^2
x_{_{\tilde{U}^3_\alpha}}(\ln x_{_{\tilde{U}^3_\alpha}}
-\ln x_{_{\chi_\beta}})}{2(x_{_{\chi_\beta}}-x_{_{\tilde{U}^3_\alpha}})^4}
\nonumber \\
&&\hspace{2.cm}+\frac{2x_{_{\chi_\beta}}^2+5x_{_{\chi_\beta}}
x_{_{\tilde{U}^3_\alpha}}
-x_{_{\tilde{U}^3_\alpha}}^2}{12(x_{_{\chi_\beta}}-
x_{_{\tilde{U}^3_\alpha}})^3}\Big]\;,
\nonumber \\
&&C_3(\mu_{_{\rm W}})=
\Big[\frac{-9x_t^2+16x_t
-4}{6(1-x_t)^4}\ln x_t+\frac{-x_t^3-11x_t^2+18x_t}{12(1-x_t)^3}
\Big]
\nonumber \\
&&\hspace{2.cm}
+\frac{1}{\tan^2\beta}\Big[\frac{(2x_tx_{_H}^3-3x_t^2x_{_H}^2)(\ln x_t
-\ln x_{_H})}{6(x_{_H}-x_t)^4}
+\frac{16x_tx_{_H}^2-29x_t^2x_{_H}+7x_t^3}{36(x_{_H}-x_t)^3}\Big]
\nonumber \\
&&\hspace{2.cm}
+\sum\limits_{\alpha,\;\beta}({\cal A}_3^{^{\alpha,\beta}})^2
\Big[\frac{x_{_{\chi_\beta}}^3
(\ln x_{_{\tilde{U}^3_\alpha}}-\ln x_{_{\chi_\beta}})}
{3(x_{_{\chi_\beta}}-x_{_{\tilde{U}^3_\alpha}})^4}
+\frac{11x_{_{\chi_\beta}}^2-7x_{_{\chi_\beta}}x_{_{\tilde{U}^3_\alpha}}+
2x_{_{\tilde{U}^3_\alpha}}^2}{18(x_{_{\chi_\beta}}
-x_{_{\tilde{U}^3_\alpha}})^3}\Big]\;,
\nonumber \\
&&C_4(\mu_{_{\rm W}})=
x_t\Big[\frac{x_t^2-x_t}
{(1-x_t)^4}\ln x_t+\frac{x_t^2-1}{2(1-x_t)^3}\Big]-\Big[\frac{x_t^2x_{_H}
(\ln x_t-\ln x_{_H})}{(x_{_H}-x_t)^3}+\frac{x_{_H}x_t+x_t^2}{2(x_{_H}-x_t)^2}\Big]
\nonumber \\
&&\hspace{2.cm}-\sum\limits_{\alpha,\;\beta}
\frac{m_{_{\chi_\beta}}}{\sqrt{2}m_{_{\rm W}}\cos\beta}
({\cal A}_3^{^{\alpha,\beta}}{\cal B}_3^{^{\alpha,\beta}})
\Big[\frac{2x_{_{\chi_\beta}}x_{_{\tilde{U}^3_\alpha}}
(\ln x_{_{\tilde{U}^3_\alpha}}-\ln x_{_{\chi_\beta}})}{(x_{_{\chi_\beta}}
-x_{_{\tilde{U}^3_\alpha}})^3}
+\frac{x_{_{\chi_\beta}}+x_{_{\tilde{U}^3_\alpha}}}{(x_{_{\chi_\beta}}
-x_{_{\tilde{U}^3_\alpha}})^2}\Big]
\;,\nonumber \\
&&C_5(\mu_{_{\rm W}})=x_t\Big[\frac{
\ln x_t}{2(1-x_t)
^3}+\frac{3-x_t}{4(1-x_t)^2}\Big]+\Big[\frac{x_tx_{_H}^2(\ln x_t-\ln x_{_H})}
{2(x_{_H}-x_t)^3}+\frac{3x_{_H}x_t-x_t^2}{4(x_{_H}-x_t)^2}\Big]
\nonumber \\
&&\hspace{2.cm}-\sum\limits_{\alpha,\;\beta}
\frac{m_{_{\chi_\beta}}}{\sqrt{2}m_{_{\rm W}}\cos\beta}
({\cal A}_3^{^{\alpha,\beta}}{\cal B}_3^{^{\alpha,\beta}})
\Big[\frac{x_{_{\chi_\beta}}x_{_{\tilde{U}^3_\alpha}}
(\ln x_{_{\tilde{U}_\alpha^3}}-\ln x_{_{\chi_\beta}})}{(x_{_{\chi_\beta}}-
x_{_{\tilde{U}^3_\alpha}})^3}
+\frac{x_{_{\chi_\beta}}+x_{_{\tilde{U}^3_\alpha}}}{2(x_{_{\chi_\beta}}
-x_{_{\tilde{U}^3_\alpha}})^2}\Big]\;,
\label{bsgwil}
\end{eqnarray}
with
\begin{eqnarray}
&&{\cal A}_i^{^{\alpha,\beta}}=-{\cal
Z}_{_{\tilde{U}^i}}^{^{1,\alpha}}{\cal Z}_{_{1,\beta}}^+
+\frac{m_{_{u_i}}}{\sqrt{2}m_{_{\rm W}}\sin\beta}{\cal
Z}_{_{\tilde{U}^i}}^{^{2,\alpha}}{\cal Z}_{_{2,\beta}}^+\;,
\nonumber \\
&&{\cal B}_i^{^{\alpha,\beta}}=-{\cal
Z}_{_{\tilde{U}^i}}^{^{1,\alpha}}{\cal Z}_{_{2,\beta}}^-\;,
\label{csusy}
\end{eqnarray}
and the mixing matrices ${\cal Z}_{_{\tilde{U}^i}}
{\cal Z}_{_{2,\beta}}^\pm$ are given in Eqs.\ref{zpm},
\ref{zud}.
The first terms in the above expressions are
the SM contributions\cite{Grigjanis}
and the second terms are the charged Higgs contributions. The supersymmetric
corrections exist in the third terms.

\begin{center}
\begin{figure}
\setlength{\unitlength}{1mm}
\begin{picture}(230,200)(55,90)
\put(50,30){\includegraphics{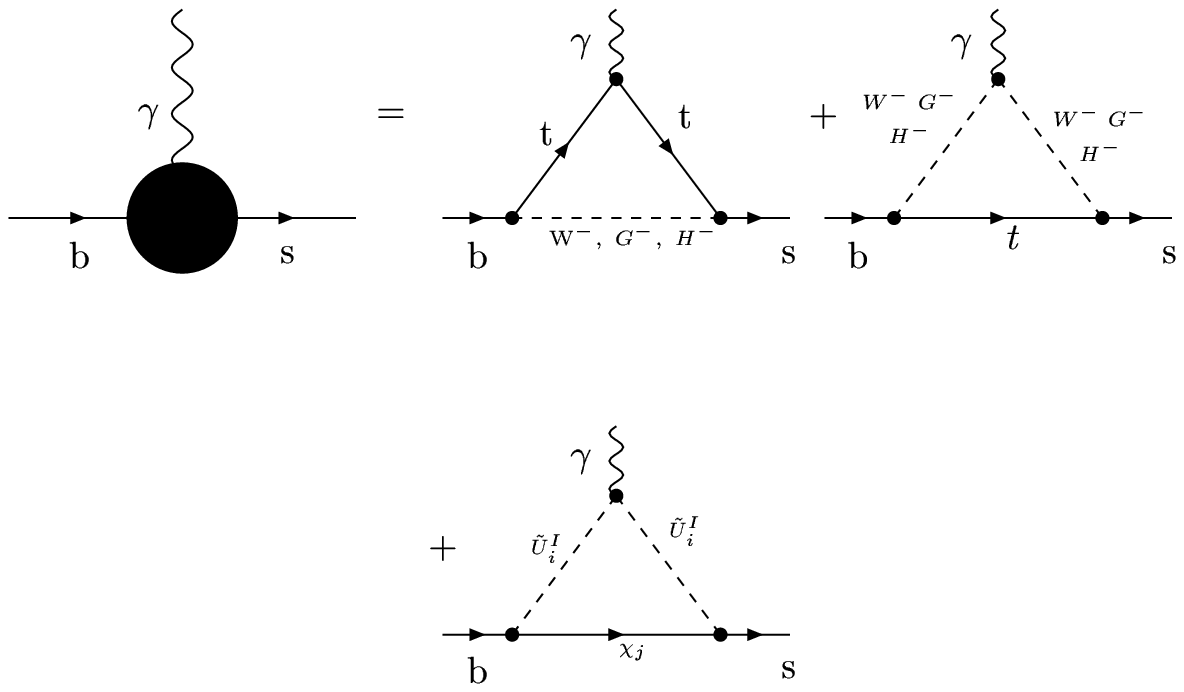}}
\end{picture}
\caption[]{The one-loop diagrams for $b\rightarrow s\gamma$
in the SUSY model with  minimal flavor violation }
\label{fig3}
\end{figure}
\end{center}

For the vertex $\bar{s}b\gamma$, the Feynman diagrams are drawn in
Fig.\ref{fig3}.

With all unrenormalized quantities the Ward-Takahashi identity
for the $\bar{s}b\gamma$ vertex is in form:
\begin{eqnarray}
&&q^\rho \Gamma_\rho^{^{b\rightarrow sg}}(p,q)=
-\frac{1}{3}e\Big[\Sigma(p+q)-\Sigma(p)\Big]\;.
\label{ward0g}
\end{eqnarray}

To renormalize the $\bar{s}b\gamma$ vertex, we demand that the Ward-Takahashi
identity be preserved for the renormalized vertex $\hat{\Gamma}
_\rho^{^{b\rightarrow s\gamma}}$\cite{Chia1},
\begin{eqnarray}
&&q^\rho \hat{\Gamma}_\rho^{^{b\rightarrow sg}}(p,q)=
-\frac{1}{3}e\Big[\hat{\Sigma}(p+q)-\hat{\Sigma}(p)\Big]\;.
\label{wardrg}
\end{eqnarray}

The other steps are similar to those applied in the
calculation for the $\bar{s}bg$ vertex. The result is written as
\begin{equation}
{\cal L}_{b\rightarrow s\gamma}=
\frac{4G_F}{\sqrt{2}}V_{tb}V_{ts}^*\Big\{C_1(\mu_{_{\rm W}}){\cal O}_1
+C_4(\mu_{_{\rm W}}){\cal O}_4+C_6(\mu_{_{\rm W}}){\cal O}_6
+C_7(\mu_{_{\rm W}}){\cal O}_7+C_8(\mu_{_{\rm W}}){\cal O}_8\Big\}
\label{bsgamma}
\end{equation}
with
\begin{eqnarray}
&&C_6(\mu_{_{\rm W}})
=x_t\Big[\frac{18x_t^2-11x_t-1}{8(1-x_t)^3}+\frac{15x_t^2-16x_t
+4}{4(1-x_t)^4}\ln x_t\Big]+\frac{x_t}{\tan^2\beta}\Big[
\frac{4x_t^2+x_tx_{_H}+25x_{_H}^2}{72(x_t-x_{_H})^3}
\nonumber \\
&&\hspace{2.cm}
-\frac{
(3x_t^2x_{_H}+2x_tx_{_H}^2)(\ln x_t-\ln x_{_H})}{12(x_t-x_{_H})^4}\Big]
+\sum\limits_{\alpha,\;\beta}({\cal A}_3^{^{\alpha,\beta}})^2
\Big[-\frac{8x_{_{\tilde{U}_\alpha^3}}^2
+5x_{_{\tilde{U}_\alpha^3}}x_{_{\chi_\beta}}-7x_{_{\chi_\beta}}^2}
{36(x_{_{\tilde{U}_\alpha^3}}-x_{_{\chi_\beta}})^2}
\nonumber \\
&&\hspace{2.cm}
+\frac{(3x_{_{\tilde{U}_\alpha^3}}^2x_{_{\chi_\beta}}
-2x_{_{\tilde{U}_\alpha^3}}x_{_{\chi_\beta}}^2)(\ln x_{_{\tilde{U}_\alpha^3}}
-\ln x_{_{\chi_\beta}})}{6(x_{_{\tilde{U}_\alpha^3}}-x_{_{\chi_\beta}})^4}\Big]
\;,\nonumber \\
&&C_7(\mu_{_{\rm W}})
=\Big[\frac{-19x_t^3+25x_t^2}{12(1-x_t)^3}+\frac{3x_t^4-30x_t^3
+54x_t^2-32x_t+8}{6(1-x_t)^4}\ln x_t\Big]
\nonumber \\
&&\hspace{2.cm}
+\frac{x_t}{\tan^2\beta}\Big[-\frac{19x_t^2+109x_tx_{_H}-98x_{_H}^2}{
108(x_t-x_{_H})^3}
+\frac{(3x_t^3-9x_t^2x_{_H}-4x_{_H}^3)(\ln x_t-\ln x_{_H})}
{36(x_t-x_{_H})^4}\Big]
\nonumber \\
&&\hspace{2.cm}+\sum\limits_{\alpha,\;\beta}({\cal A}_3^{^{\alpha,\beta}})^2
\Big[\frac{52x_{_{\tilde{U}_\alpha^3}}^2
-101x_{_{\tilde{U}_\alpha^3}}x_{_{\chi_\beta}}+43x_{_{\chi_\beta}}^2}
{54(x_{_{\tilde{U}_\alpha^3}}-x_{_{\chi_\beta}})^3}
\nonumber \\
&&\hspace{2.cm}
-\frac{(6x_{_{\tilde{U}_\alpha^3}}^3-27x_{_{\tilde{U}_\alpha^3}}^2
x_{_{\chi_\beta}}+12x_{_{\tilde{U}_\alpha^3}}x_{_{\chi_\beta}}^2
+2x_{_{\chi_\beta}}^3)(\ln x_{_{\tilde{U}_\alpha^3}}-\ln x_{_{\chi_\beta}})}
{9(x_{_{\tilde{U}_\alpha^3}}-x_{_{\chi_\beta}})^4}\Big]
\;,\nonumber \\
&&C_8(\mu_{_{\rm W}})
=x_t\Big[\frac{-5x_t^2+8x_t-3}{4(1-x_t)^3}+\frac{3x_t-2}{2(1-x_t)^3}
\ln x_t\Big]
\nonumber \\
&&\hspace{2.cm}
-x_t\Big[\frac{1}{2(x_t-x_{_H})}
+\frac{(x_{_H}^2+x_tx_{_H})(\ln x_t-\ln x_{_H})}{2(x_t-x_{_H})^3}\Big]
\nonumber \\
&&\hspace{2.cm}
+\sum\limits_{\alpha,\;\beta}
\frac{m_{_{\chi_\beta}}}{\sqrt{2}m_{_{\rm W}}\cos\beta}
({\cal A}_3^{^{\alpha,\beta}}{\cal B}_3^{^{\alpha,\beta}})
\Big[-\frac{7x_{_{\tilde{U}_\alpha^3}}-5x_{_{\chi_\beta}}}
{6(x_{_{\tilde{U}_\alpha^3}}-x_{_{\chi_\beta}})^2}\nonumber \\
&&\hspace{2.cm}
+\frac{(3x_{_{\tilde{U}_\alpha^3}}^2+2x_{_{\tilde{U}_\alpha^3}}
x_{_{\chi_\beta}})(\ln x_{_{\tilde{U}_\alpha^3}}-\ln x_{_{\chi_\beta}})}
{3(x_{_{\tilde{U}_\alpha^3}}-x_{_{\chi_\beta}})^3}\Big]
\;.   \nonumber \\
\label{bsgammawil}
\end{eqnarray}

\section{Numerical results}

In this section, we present our numerical results of the Wilson
coefficients in the mSUGRA model.
As we have mentioned before, the model is fully specified by five
parameters $$m_0,\;m_\frac{1}{2},\;A_0,\;\tan\beta,\;sgn(\mu).$$
Here $m_0,\;m_{\frac{1}{2}},\;$ and $A_0$ are the universal
scalar quark mass, gaugino mass and trilinear scalar coupling. They are
assumed to arise through supersymmetry breaking in a hidden-sector
at the GUT scale $\mu_{GUT}\simeq 2\times 10^{16}$GeV. In our
numerical calculation, to maintain consistency of the theory and
the up-to-date
experimental observation, when we obtain the numerical value of the Higgs mass
in the mSUGRA model with the five parameters, and
we include all one-loop effects in the
Higgs potential\cite{Pierce}. Moreover  we also employ the two-loop
RGEs\cite{2loop} with one-loop threshold
corrections\cite{Pierce,Bagger1} as the energy scale runs down from the mSUGRA
scale to the lower weak scale.

For the SM parameters, we have $m_b=5{\rm GeV},\; m_t=174{\rm GeV}
,\; m_{\rm W}=80.23{\rm GeV},\;\alpha_e(m_{\rm W})=\frac{1}{128},\;
\alpha_s(m_{\rm W})=0.12$ at the weak scale.
In our later calculations, we always set $A_0=0,\; sgn(\mu)=-$.
Taking above values, we find that
the Standard Model prediction for
the Wilson Coefficients are  $C_{2_{SM}}(m_{_{\rm W}})=0.315
,\;C_{3_{SM}}(m_{_{\rm W}})=0.256,\;C_{5_{SM}}(m_{_{\rm W}})=-0.218,
\;C_{6_{SM}}(m_{_{\rm W}})=-1.477
,\;C_{7_{SM}}(m_{_{\rm W}})=1.511,\;C_{8_{SM}}(m_{_{\rm W}})=0.891$.
Then with the aforementioned inputs of the five parameter
we evaluate the supersymmetric corrections to those Wilson
coefficients $C_i(m_{_{\rm W}})\;(i=2,\;3,\;5,\;\cdots,8)$.

Even though other Wilson
coefficients also get nonzero contributions from the supersymmetric
sector, our discussions mainly focus on
$C_{_{2,3,5,6,7,8}}(m_{_{\rm W}})$ because they play more
significant roles in low energy  phenomenology.
Moreover, we will illustrate their dependence on  the
supersymmetric parameters through the attached figures.

In Fig.\ref{fig4}, we plot $C_2(m_{_{\rm W}}),\;
C_3(m_{_{\rm W}}),\;C_5(m_{_{\rm W}})$ versus
$m_{\frac{1}{2}}$ with $m_0=100{\rm GeV},\; A_0=0,\;
sgn(\mu)=-$ and $\tan\beta=2,\;20$. Setting $m_{1\over 2}=100{\rm GeV},\;
A_0=0,\; sgn(\mu)=-$ and $\tan\beta=2,\;20$,
the dependence of $C_2(m_{_{\rm W}}),\;
C_3(m_{_{\rm W}}),\;C_5(m_{_{\rm W}})$ on $m_0$ is
plotted in Fig.\ref{fig5}. From Fig.\ref{fig4}, we find that the
supersymmetric contributions make the Wilson
coefficients at the weak scale deviate from the SM predictions obviously when
$m_{1\over 2}\leq 800{\rm GeV}$; when $m_{1\over 2}$ further
increases, the new physics contributions gradually become immaterial.
A similar situation exists in Fig.\ref{fig5},  the Wilson
coefficients tend to the SM prediction values as
$m_0$ increases. For the Wilson coefficients of $b\rightarrow
s\gamma$, we plot  $C_6(m_{_{\rm W}}),\;
C_7(m_{_{\rm W}}),\;C_8(m_{_{\rm W}})$ versus
$m_{\frac{1}{2}}$ with $m_0=100{\rm GeV},\; A_0=0,\;
sgn(\mu)=-$ and $\tan\beta=2,\;20$ in Fig.\ref{fig6}.
Setting $m_{1\over 2}=100{\rm GeV},\;
A_0=0,\; sgn(\mu)=-$ and $\tan\beta=2,\;20$,
dependence of $C_2(m_{_{\rm W}}),\;
C_3(m_{_{\rm W}}),\;C_5(m_{_{\rm W}})$ on $m_0$ is
plotted in Fig.\ref{fig7}. The trend of changes of those coefficients
with respect to
parameters $m_{1\over 2},\; m_0$ is similar to that in the $b\rightarrow sg$
case.

When the effective Lagrangian is applying at the hadronic scale, we
should evolve those Wilson coefficients from the weak scale down
to the hadronic scale. The running depends on the anomalous dimension
matrix of concerned operators\cite{Buras1}.
The coefficients $C_i(m_{_{\rm W}})$ obtained at the weak scale $M_W$ are
regarded as the initial conditions for the differential RGEs.

\section{Discussions}

In this work, we discuss contributions to the
effective Lagrangian for $b\rightarrow sg$ and $b\rightarrow s\gamma$
from the SUSY sector in the mSUGRA model. As many authors
suggested, if the masses of the lightest SUSY particles are close
to the electroweak energy scale, the contribution from the SUSY
sector to the Wilson coefficients of the induced operators is
comparable with that from SM.

Our numerical results indicate that within a reasonable mSUGRA
parameter range, the SUSY contribution to $C_5(m_{_{\rm W}})$
can enlarge the SM prediction by about 90\%, and to the other coefficients,
the SUSY contributions are not smaller than 30\% of that of SM.

If the masses of the SUSY particles are larger, the SUSY
contributions to the effective Lagrangian would become weaker. Then
as the SUSY particles are very heavy, the main contribution to the
effective Lagrangian uniquely comes from the standard model.

As well known, the QCD correction to the vertices $\bar sbg$ and
$\bar sb\gamma$ is important and for practical application of the
effective Lagrangian to evaluate the
physical processes, say B decays, one needs to run down
the coefficients from the weak energy scale to the hadronic scale,
i.e. $M_B\sim 5$GeV$\ll \mu_W$, in terms of the RGEs\cite{fengtf}.

In this work, we adopt the non-linear $R_{\xi}$ gauge. The
advantage is that the Ward-Takahashi identity holds at the
one-loop level no matter for the unrenormalized or renormalized
quantities. This advantage would be more obvious as we go on doing
the two-loop calculations.

Our numerical results also show that as all SUSY particles become
very heavy, the values of all coefficients tend to that
determined by the SM sector which
is consistent with the results obtained before \cite{Chia1}.

\vspace{1.0cm}
\noindent {\Large{\bf Acknowledgements}}

This work is partially supported by the National Natural Science
Foundation of China.

\vspace{1cm}
\vspace{1.0cm}
\noindent {\Large{\bf Appendix}}
\appendix
\section{The expressions for the form factors \label{app1}}

The form factors in self-energy is given as
\begin{eqnarray}
&&A_0(x_i,x_{_H},x_{_{\tilde{U}_\alpha^i}},x_{_{\chi_\beta}})=
(1+\frac{x_i}{2})\Big[\Delta +\frac{1}{2}
+\ln x_\mu+\frac{x_i}{x_i-1}-\frac{x_i^2\ln x_i}{(x_i-1)^2}\Big]
\nonumber \\
&&\hspace{3.8cm}+\frac{x_i}{2\tan^2\beta}\Big[\Delta+\frac{1}{2}
+\ln x_\mu+\frac{x_i}{x_i-x_{_H}}-\frac{x_i^2\ln x_i}{(x_i-x_{_H})^2}
+\frac{(2x_{_H}x_i-x_{_H}^2)\ln x_{_H}}{(x_i-x_{_H})^2}\Big]
\nonumber \\
&&\hspace{3.8cm}
+\sum\limits_{\alpha,\;\beta}({\cal A}_i^{^{\alpha,\beta}})^2
\Big[\Delta+\frac{3}{2}+\ln x_\mu-\frac{x_{_{\tilde{U}_\alpha^i}}}
{x_{_{\tilde{U}_\alpha^i}}-x_{_{\chi_\beta}}}+\frac{x_{_{\tilde{U}_\alpha^i}}
(2x_{_{\chi_\beta}}-x_{_{\tilde{U}_\alpha^i}})\ln x_{_{\tilde{U}_\alpha^i}}}
{(x_{_{\tilde{U}_\alpha^i}}-x_{_{\chi_\beta}})^2}
-\frac{x_{_{\chi_\beta}}^2\ln x_{_{\chi_\beta}}}
{(x_{_{\tilde{U}_\alpha^i}}-x_{_{\chi_\beta}})^2}\Big]\;,
\nonumber \\
&&A_1(x_i,x_{_H},x_{_{\tilde{U}_\alpha^i}},x_{_{\chi_\beta}})=
(1+\frac{x_i}{2})\Big[\frac{2x_i^2+5x_i-1}{3(x_i-1)^3}
-\frac{2x_i^2\ln x_i}{(x_i-1)^4}\Big]\nonumber \\
&&\hspace{3.8cm}
+\frac{x_i}{2\tan^2\beta}\Big[\frac{2x_i^2+5x_ix_{_H}-x_{_H}^2}{3(x_i-x_{_H})^3}
-\frac{2x_i^2x_{_H}(\ln x_i-\ln x_{_H})}{(x_i-x_{_H})^4}\Big]
\nonumber\\
&&\hspace{3.8cm}+\sum\limits_{\alpha,\;\beta}({\cal A}_i^{^{\alpha,\beta}})^2
\Big[\frac{x_{_{\tilde{U}_\alpha^i}}^2
-5x_{_{\tilde{U}_\alpha^i}}x_{_{\chi_\beta}}
-2x_{_{\chi_\beta}}^2}{3(x_{_{\tilde{U}_\alpha^i}}-x_{_{\chi_\beta}})^3}
+\frac{2x_{_{\tilde{U}_\alpha^i}}x_{_{\chi_\beta}}(\ln x_{_{\tilde{U}_\alpha^i}}
-\ln
x_{_{\chi_\beta}})}{(x_{_{\tilde{U}_\alpha^i}}-x_{_{\chi_\beta}})^4}\Big]
\;, \nonumber \\
&&B_0(x_i,x_{_H},x_{_{\tilde{U}_\alpha^i}},x_{_{\chi_\beta}})=
\frac{x_i^2\ln x_i}{x_i-1}
-\frac{x_i^2\ln x_i
-x_ix_{_H}\ln x_{_H}}{x_i-x_{_H}}
\nonumber \\
&&\hspace{3.8cm}+2\sum\limits_{\alpha,\;\beta}
\frac{m_{_{\chi_\beta}}}{\sqrt{2}m_{_{\rm W}}\cos\beta}
({\cal A}_3^{^{\alpha,\beta}}{\cal B}_3^{^{\alpha,\beta}})
\Big[\Delta+1+\ln x_\mu-\frac{x_{_{\tilde{U}_\alpha^i}}
\ln x_{_{\tilde{U}_\alpha^i}}-x_{_{\chi_\beta}}\ln x_{_{\chi_\beta}}}
{x_{_{\tilde{U}_\alpha^i}}-x_{_{\chi_\beta}}}\Big]\;,
\nonumber \\
&&B_1(x_i,x_{_H},x_{_{\tilde{U}_\alpha^i}},x_{_{\chi_\beta}})=
-x_i\Big[\frac{x_i+1}{2(x_i-1)^2}-\frac{x_i\ln x_i}
{(x_i-1)^3}\Big]\nonumber \\
&&\hspace{3.8cm}
+x_i\Big[\frac{x_i+x_{_H}}{2(x_i-x_{_H})^2}
-\frac{x_ix_{_H}(\ln x_i-\ln x_{_H})}{(x_i-x_{_H})^3}\Big]
\nonumber \\
&&\hspace{3.8cm}
+\sum\limits_{\alpha,\;\beta}
\frac{m_{_{\chi_\beta}}}{\sqrt{2}m_{_{\rm W}}\cos\beta}
({\cal A}_3^{^{\alpha,\beta}}{\cal B}_3^{^{\alpha,\beta}})
\Big[\frac{x_{_{\tilde{U}_\alpha^i}}+x_{_{\chi_\beta}}}
{(x_{_{\tilde{U}_\alpha^i}}-x_{_{\chi_\beta}})^2}
-\frac{2x_{_{\tilde{U}_\alpha^i}}x_{_{\chi_\beta}}(\ln x_{_{\tilde{U}_\alpha^i}}
-\ln x_{_{\chi_\beta}})}{(x_{_{\tilde{U}_\alpha^i}}-x_{_{\chi_\beta}})^3}\Big]\;,
\nonumber \\
&&C_0(x_i,x_{_H},x_{_{\tilde{U}_\alpha^i}},x_{_{\chi_\beta}})=
\frac{1}{2}\Big[\Delta+\frac{1}{2}+\ln x_\mu
+\frac{x_i}{x_i-1}-\frac{x_i^2\ln x_i}{(x_i-1)^2}\Big]
\nonumber\\
&&\hspace{3.8cm}+\frac{\tan^2\beta}{2}\Big[\Delta+\frac{1}{2}
+\ln x_\mu+\frac{x_i}{x_i-x_{_H}}-\frac{x_i^2\ln x_i}{(x_i-x_{_H})^2}
+\frac{(2x_{_H}x_i-x_{_H}^2)\ln x_{_H}}{(x_i-x_{_H})^2}\Big]
\nonumber \\
&&\hspace{3.8cm}+\sum\limits_{\alpha,\;\beta}
\frac{({\cal B}_3^{^{\alpha,\beta}})^2}{2\cos^2\beta}
\Big[\Delta+\frac{3}{2}+\ln x_\mu-\frac{x_{_{\tilde{U}_\alpha^i}}}
{x_{_{\tilde{U}_\alpha^i}}-x_{_{\chi_\beta}}}+\frac{x_{_{\tilde{U}_\alpha^i}}
(2x_{_{\chi_\beta}}-x_{_{\tilde{U}_\alpha^i}})\ln x_{_{\tilde{U}_\alpha^i}}}
{(x_{_{\tilde{U}_\alpha^i}}-x_{_{\chi_\beta}})^2}
\nonumber \\
&&\hspace{3.8cm}-\frac{x_{_{\chi_\beta}}^2\ln x_{_{\chi_\beta}}}
{(x_{_{\tilde{U}_\alpha^i}}-x_{_{\chi_\beta}})^2}\Big]\;.
\label{exp1}
\end{eqnarray}
The expression for $F_i(x_i,x_{_H},x_{_{\tilde{U}_\alpha^i}},x_{_{\chi_\beta}})
\;(i=1,\;\cdots,\;5)$ is written as
\begin{eqnarray}
&&F_1(x_i,x_{_H},x_{_{\tilde{U}_\alpha^i}},x_{_{\chi_\beta}})=
(1+\frac{x_i}{2})\Big[\frac{5x_i^2-22x_i+5}{18(x_i-1)^3}+\frac{(3x_i-1)\ln x_i}
{3(x_i-1)^4}\Big]
\nonumber \\
&&\hspace{3.8cm}+\frac{1}{\tan^2\beta}\Big[\frac{x_i(5x_i^2-22x_ix_{_H}+5x_{
_H}^2)}{36(x_i-x_{_H})^3}-\frac{(x_{_H}^3x_i-3x_{_H}^2x_i^2)(\ln x_i-\ln x_{_H})}
{6(x_i-x_{_H})^4}\Big]
\nonumber \\
&&\hspace{3.8cm}+\sum\limits_{\alpha,\;\beta}({\cal A}_i^{^{\alpha,\beta}})^2
\Big[\frac{x_{_{\tilde{U}_\alpha^i}}^2
-8x_{_{\tilde{U}_\alpha^i}}x_{_{\chi_\beta}}-17x_{_{\chi_\beta}}^2}
{36(x_{_{\tilde{U}_\alpha^i}}-x_{_{\chi_\beta}})^3}
+\frac{(3x_{_{\tilde{U}_\alpha^i}}x_{_{\chi_\beta}}^2+x_{_{\chi_\beta}}^3)
(\ln x_{_{\tilde{U}_\alpha^i}}-\ln x_{_{\chi_\beta}})}
{6(x_{_{\tilde{U}_\alpha^i}}-x_{_{\chi_\beta}})^4}\Big]\;,
\nonumber \\
&&F_2(x_i,x_{_H},x_{_{\tilde{U}_\alpha^i}},x_{_{\chi_\beta}})=
\Big[-\frac{x_i^3-15x_i^2-12x_i+8}{12(x_i-1)^3}-\frac{(5x_i^2-2x_i)\ln x_i}
{2(x_i-1)^4}\Big]
\nonumber \\
&&\hspace{3.8cm}+\frac{1}{\tan^2\beta}\Big[-\frac{x_i(x_i^2-5x_ix_{_H}-2x_{_H}^2)
}{12(x_i-x_{_H})^3}-\frac{x_i^2x_{_H}^2(\ln x_i-\ln x_{_H})}{2(x_i-x_{_H})^4}
\Big]
\nonumber \\
&&\hspace{3.8cm}+\sum\limits_{\alpha,\;\beta}({\cal A}_i^{^{\alpha,\beta}})^2
\Big[-\frac{x_{_{\tilde{U}_\alpha^i}}^2
-x_{_{\tilde{U}_\alpha^i}}x_{_{\chi_\beta}}-2x_{_{\chi_\beta}}^2}
{6(x_{_{\tilde{U}_\alpha^i}}-x_{_{\chi_\beta}})^3}
-\frac{x_{_{\tilde{U}_\alpha^i}}x_{_{\chi_\beta}}^2
(\ln x_{_{\tilde{U}_\alpha^i}}-\ln x_{_{\chi_\beta}})}
{(x_{_{\tilde{U}_\alpha^i}}-x_{_{\chi_\beta}})^4}\Big]\;,
\nonumber \\
&&F_3(x_i,x_{_H},x_{_{\tilde{U}_\alpha^i}},x_{_{\chi_\beta}})=
\Big[\frac{5x_i^3+3x_i^2+6x_i+4}{12(x_i-1)^3}+\frac{x_i(2x_i^2-x_i-2)\ln x_i}
{2(x_i-1)^4}\Big]
\nonumber \\
&&\hspace{3.8cm}+\frac{1}{\tan^2\beta}\Big[\frac{x_i(5x_i^2+5x_ix_{_H}-4x_{_H}^2)
}{12(x_i-x_{_H})^3}+\frac{(2x_i^3x_{_H}-x_i^2x_{_H}^2)(\ln x_i-\ln x_{_H})}
{2(x_i-x_{_H})^4}\Big]
\nonumber \\
&&\hspace{3.8cm}+\sum\limits_{\alpha,\;\beta}({\cal A}_i^{^{\alpha,\beta}})^2
\Big[-\frac{x_{_{\tilde{U}_\alpha^i}}^2
-x_{_{\tilde{U}_\alpha^i}}x_{_{\chi_\beta}}-2x_{_{\chi_\beta}}^2}
{6(x_{_{\tilde{U}_\alpha^i}}-x_{_{\chi_\beta}})^3}
-\frac{x_{_{\tilde{U}_\alpha^i}}x_{_{\chi_\beta}}^2
(\ln x_{_{\tilde{U}_\alpha^i}}-\ln x_{_{\chi_\beta}})}
{(x_{_{\tilde{U}_\alpha^i}}-x_{_{\chi_\beta}})^4}\Big]\; ,
\nonumber \\
&&F_4(x_i,x_{_H},x_{_{\tilde{U}_\alpha^i}},x_{_{\chi_\beta}})=
-(1+\frac{x_i}{2})\Big[\frac{5x_i^2-22x_i+5}{18(x_i-1)^3}+\frac{(3x_i-1)\ln x_i}
{3(x_i-1)^4}\Big]
\nonumber \\
&&\hspace{3.8cm}+\frac{1}{\tan^2\beta}\Big[-\frac{x_i(5x_i^2-22x_ix_{_H}
+5x_{_H}^2)}{36(x_i-x_{_H})^3}+\frac{(x_{_H}^3x_i-3x_{_H}^2x_i^2)
(\ln x_i-\ln x_{_H})}{6(x_i-x_{_H})^4}\Big]
\nonumber \\
\nonumber \\
&&\hspace{3.8cm}+\sum\limits_{\alpha,\;\beta}({\cal A}_i^{^{\alpha,\beta}})^2
\Big[\frac{x_{_{\tilde{U}_\alpha^i}}^2
-8x_{_{\tilde{U}_\alpha^i}}x_{_{\chi_\beta}}-17x_{_{\chi_\beta}}^2}
{36(x_{_{\tilde{U}_\alpha^i}}-x_{_{\chi_\beta}})^3}
+\frac{(3x_{_{\tilde{U}_\alpha^i}}x_{_{\chi_\beta}}^2+x_{_{\chi_\beta}}^3)
(\ln x_{_{\tilde{U}_\alpha^i}}-\ln x_{_{\chi_\beta}})}
{6(x_{_{\tilde{U}_\alpha^i}}-x_{_{\chi_\beta}})^4}\Big]\; ,
\nonumber \\
&&F_5(x_i,x_{_H},x_{_{\tilde{U}_\alpha^i}},x_{_{\chi_\beta}})=
\Big[-\frac{x_i(x_i-3)}{4(x_i-1)^2}-\frac{x_i\ln x_i}{2(x_i-1)^3}\Big]
+\Big[\frac{x_i(x_i-3x_{_H})}{4(x_i-x_{_H})^2}+
\frac{x_ix_{_H}^2(\ln x_i-\ln x_{_H})}{2(x_i-x_{_H})^3}\Big]
\nonumber \\
&&\hspace{3.8cm}+\sum\limits_{\alpha,\;\beta}
\frac{m_{_{\chi_\beta}}}{\sqrt{2}m_{_{\rm W}}\cos\beta}
({\cal A}_3^{^{\alpha,\beta}}{\cal B}_3^{^{\alpha,\beta}})
\Big[\frac{x_{_{\tilde{U}_\alpha^i}}+x_{_{\chi_\beta}}}
{2(x_{_{\tilde{U}_\alpha^i}}-x_{_{\chi_\beta}})^2}
+\frac{x_{_{\tilde{U}_\alpha^i}}x_{_{\chi_\beta}}
(-\ln x_{_{\tilde{U}_\alpha^i}}+\ln x_{_{\chi_\beta}})}
{(x_{_{\tilde{U}_\alpha^i}}-x_{_{\chi_\beta}})^3}\Big]\;.
\end{eqnarray}


\begin{center}
\begin{figure}
\setlength{\unitlength}{1mm}
\begin{picture}(230,200)(55,90)
\put(50,80){\includegraphics{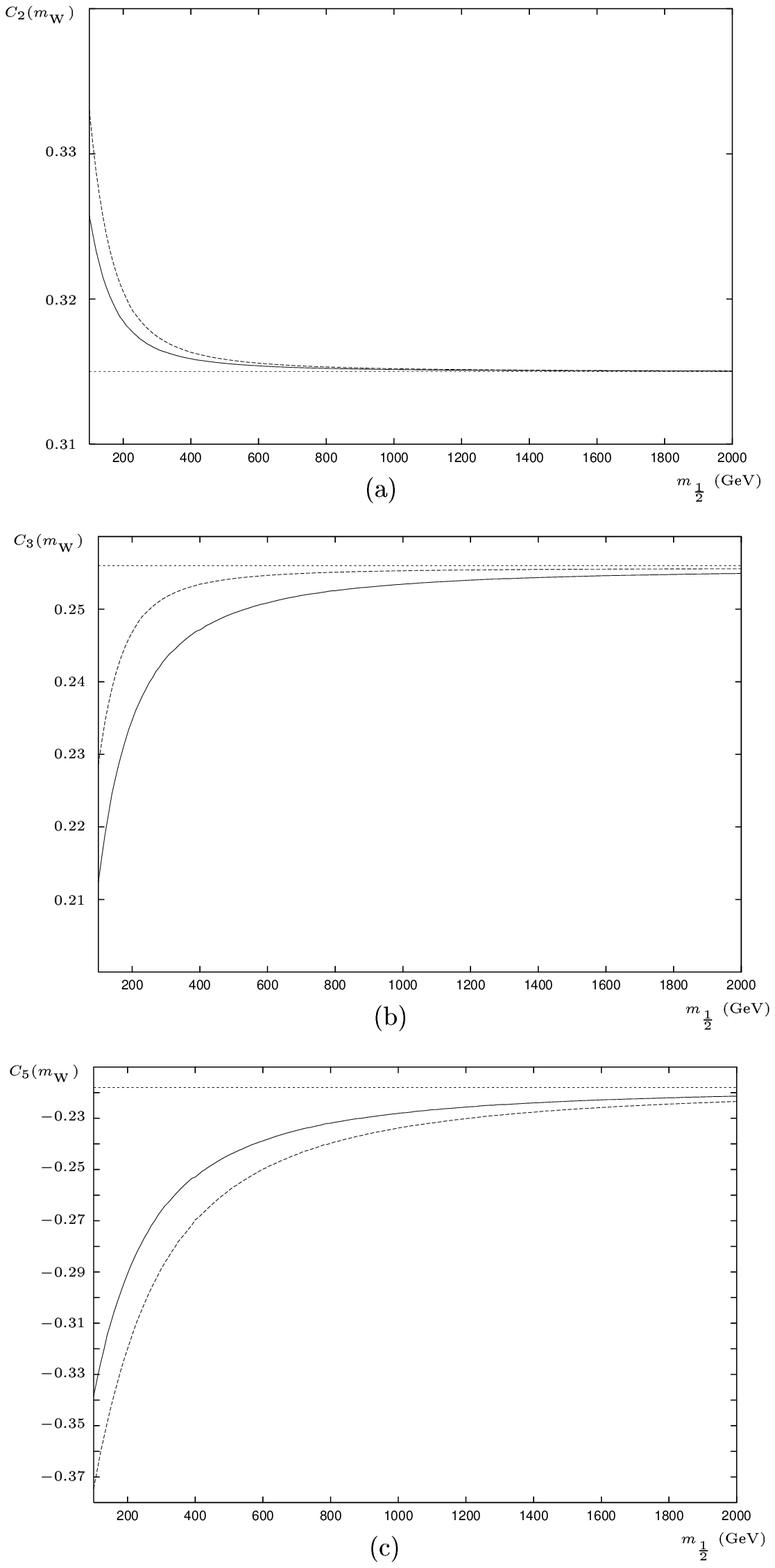}}
\end{picture}
\caption[]{The Wilson coefficients at the weak scale for $b\rightarrow sg$
in mSUGRA versus $m_\frac{1}{2}$ with $A_0=0,\; sgn(\mu)=-$ and (i)
solid-line: $\tan\beta=2$; (ii) dash-line: $\tan\beta=20$;
(iii) dot-line: the SM prediction value.}
\label{fig4}
\end{figure}

\begin{figure}
\setlength{\unitlength}{1mm}
\begin{picture}(230,200)(55,90)
\put(50,80){\includegraphics{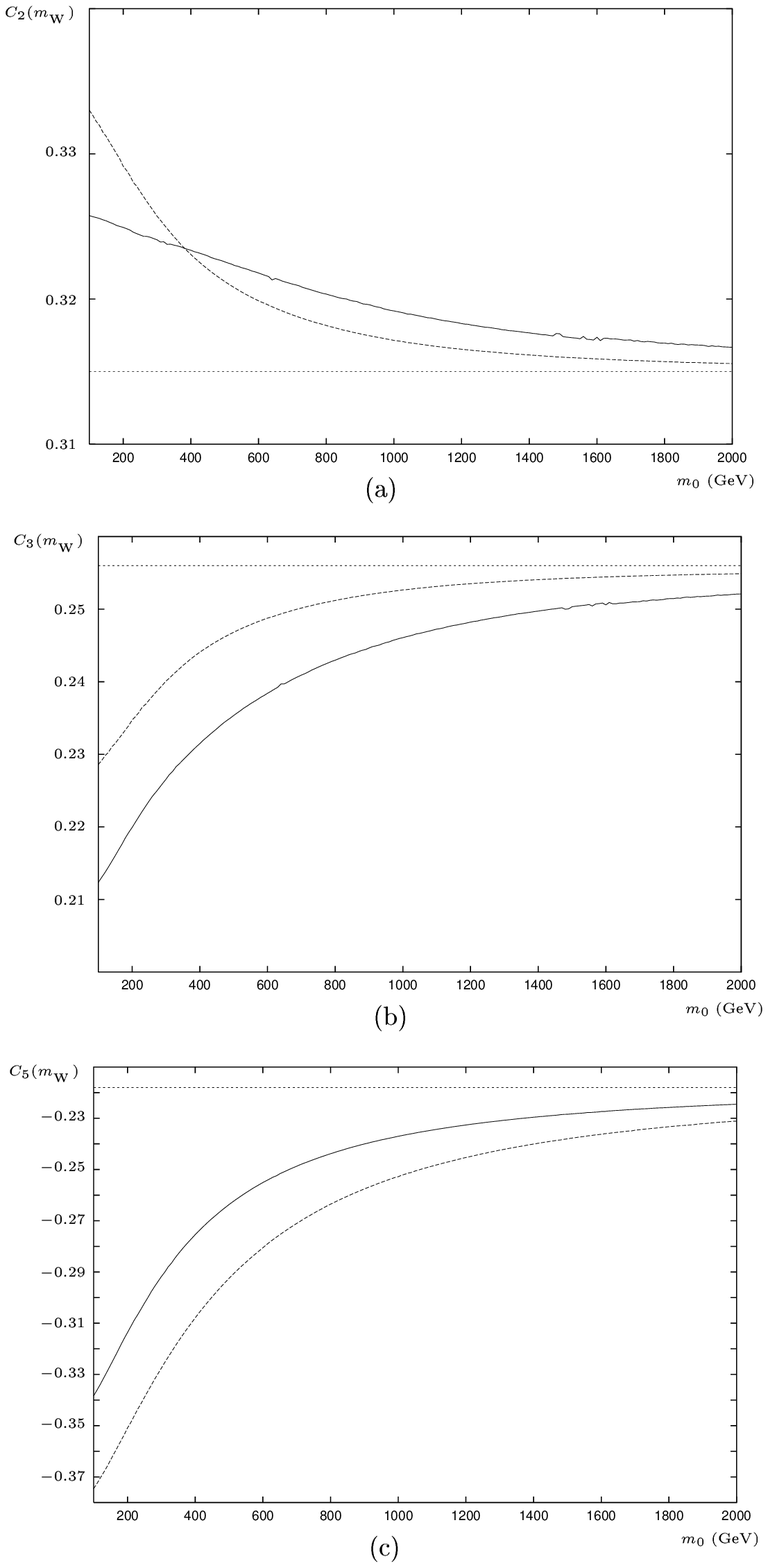}}
\end{picture}
\caption[]{The Wilson coefficients at the weak scale for $b\rightarrow sg$
in mSUGRA versus $m_0^2$ with $A_0=0,\; sgn(\mu)=-$ and (i)
solid-line: $\tan\beta=2$; (ii) dash-line: $\tan\beta=20$;
(iii) dot-line: the SM prediction value.}
\label{fig5}
\end{figure}

\begin{figure}
\setlength{\unitlength}{1mm}
\begin{picture}(230,200)(55,90)
\put(50,80){\includegraphics{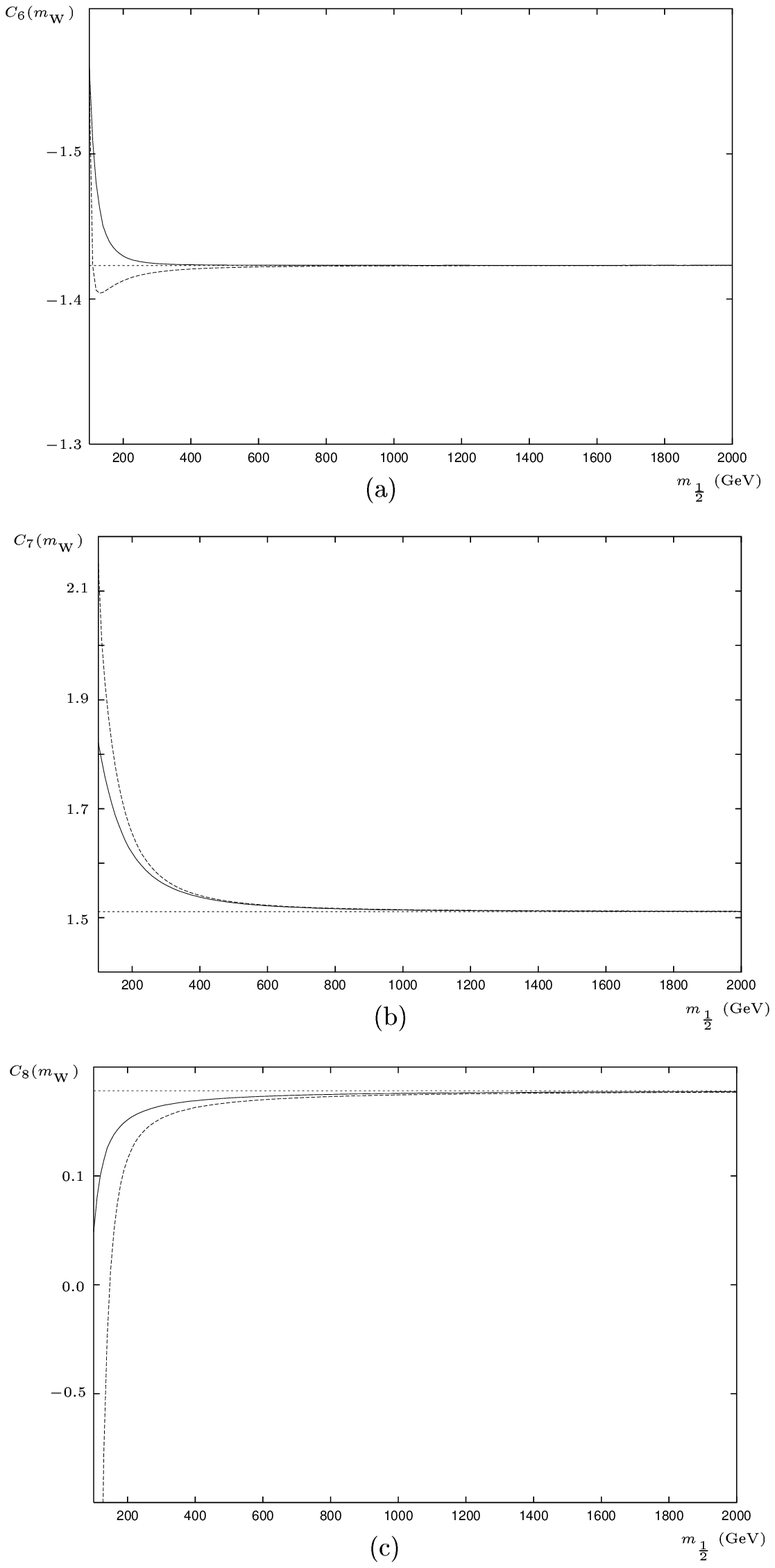}}
\end{picture}
\caption[]{The Wilson coefficients at the weak scale for $b\rightarrow s\gamma$
in mSUGRA versus $m_\frac{1}{2}$ with $A_0=0,\; sgn(\mu)=-$ and (i)
solid-line: $\tan\beta=2$; (ii) dash-line: $\tan\beta=20$;
(iii) dot-line: the SM prediction value.}
\label{fig6}
\end{figure}

\begin{figure}
\setlength{\unitlength}{1mm}
\begin{picture}(230,200)(55,90)
\put(50,80){\includegraphics{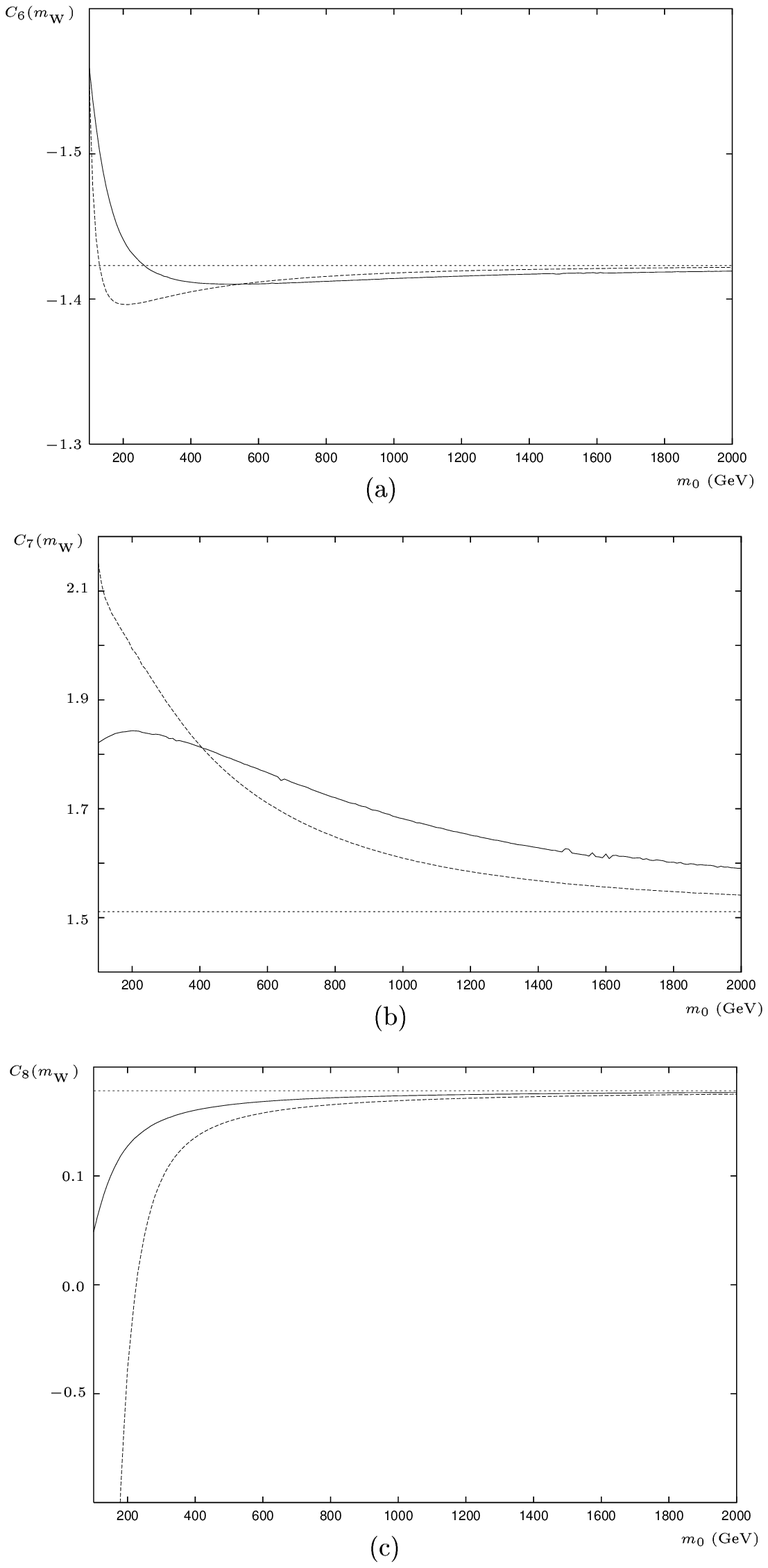}}
\end{picture}
\caption[]{The Wilson coefficients at the weak scale for $b\rightarrow s\gamma$
in mSUGRA versus $m_0^2$ with $A_0=0,\; sgn(\mu)=-$ and (i)
solid-line: $\tan\beta=2$; (ii) dash-line: $\tan\beta=20$;
(iii) dot-line: the SM prediction value.}
\label{fig7}
\end{figure}
\end{center}
\end{document}